\documentclass[twocolumn]{aastex62}

\usepackage[utf8]{inputenc}
\usepackage{amsmath, amsthm, amssymb}
\usepackage{savesym}
\savesymbol{tablenum}
\usepackage{siunitx}
\restoresymbol{SIX}{tablenum}
\usepackage[italicdiff]{physics}
\usepackage{graphicx}
\usepackage[caption=false]{subfig}
\usepackage{cleveref}
\usepackage{scrextend}
\usepackage{tabulary,booktabs,longtable}

\DeclareSIUnit{\year}{yr}
\DeclareSIUnit{\parsec}{pc}
\DeclareSIUnit{\Msun}{M_{\odot}}

\defcitealias{mingarelli_local_2017a}{M17}
\defcitealias{goulding_discovery_2019}{G19}
\defcitealias{sesana_systematic_2013a}{S13}
\defcitealias{hopkins_observational_2007}{H7}

\shorttitle{Quasar-based SMBHB population models: implications for the GWB}
\shortauthors{Casey-Clyde, Mingarelli, Greene et al.}

\begin{document}

    \title{A quasar-based supermassive black hole binary population model: implications for the gravitational-wave background}

    \correspondingauthor{Andrew Casey-Clyde}
    \email{andrew.casey-clyde@uconn.edu}

\author[0000-0002-5557-4007]{J. Andrew Casey-Clyde}
\affil{Department of Physics, University of Connecticut, 196 Auditorium Road, U-3046, Storrs, CT 06269-3046, USA}

\author[0000-0002-4307-1322]{Chiara M.  F.  Mingarelli}
\affil{Department of Physics, University of Connecticut, 196 Auditorium Road, U-3046, Storrs, CT 06269-3046, USA}
\affil{Center for Computational Astrophysics, Flatiron Institute, 162 Fifth Ave, New York, NY, 10010, USA}

\author[0000-0002-5612-3427]{Jenny E. Greene}
\affil{Department of Astrophysical Sciences, Princeton University, Princeton, NJ 08544, USA}

\author[0000-0002-9910-6782]{Kris Pardo}
\affil{Jet Propulsion Laboratory, California Institute of Technology, Pasadena, CA 91125, USA}

\author{Morgan Na\~{n}ez}
\affil{Department of Astronomy, University of California Berkeley, Berkeley CA 94720, USA}

\author{Andy D. Goulding}
\affil{Department of Astrophysical Sciences, Princeton University, Princeton, NJ 08544, USA}

    \begin{abstract}
        The nanohertz gravitational wave background (GWB) is believed to be dominated by GW emission from supermassive black hole binaries (SMBHBs).
         Observations of several dual active galactic nuclei (AGN) strongly suggest a link between AGN and SMBHBs, given that these dual AGN systems will eventually form bound binary pairs.
        Here we develop an exploratory SMBHB population model based on empirically constrained quasar populations, allowing us to decompose the GWB amplitude into an underlying distribution of SMBH masses, SMBHB number density, and volume enclosing the GWB. Our approach also allows us to self-consistently predict the number of local SMBHB systems from the GWB amplitude. Interestingly, we find the local number density of SMBHBs implied by the common-process signal in the NANOGrav 12.5-yr dataset to be roughly five times larger than previously predicted by other models.
        We also find that at most $\sim 25 \%$ of SMBHBs can be associated with quasars.
        Furthermore, our quasar-based approach predicts $\gtrsim 95\%$ of the GWB signal comes from $z \lesssim 2.5$, and that SMBHBs contributing to the GWB have masses $\gtrsim 10^8 M_\odot$.
        We also explore how different empirical galaxy-black hole scaling relations affect the local number density of GW sources, and find that relations predicting more massive black holes decrease the local number density of SMBHBs.
        Overall, our results point to the important role that a measurement of the GWB will play in directly constraining the cosmic population of SMBHBs, as well as their connections to quasars and galaxy mergers.
    \end{abstract}


    \section{Introduction}
        Gravitational waves (GWs) from the inspiral and merger of supermassive black hole binaries (SMBHBs) should be the most powerful GWs in the Universe~\citep{volonteri_assembly_2003,burke-spolaor_astrophysics_2019, mingarelli_probing_2019}. Their incoherent superposition forms a GW background (GWB), which may already be manifesting in pulsar timing array (PTA) datasets, including the NANOGrav 12.5-year dataset~\citep{arzoumanian_nanograv_2020a} and the second data release from the Parkes Pulsar Timing Array~\citep[PPTA,][]{goncharov_evidence_2021a}. The amplitude of this GWB is rich in information about the underlying SMBH population which is difficult to access any other way: it depends on the mass of the contributing black holes, the volume enclosing the background, and the distribution of SMBHBs throughout that volume~\citep[see e.g.][S13 hereafter]{phinney_practical_2001a, sesana_systematic_2013a}.
        
        In hierarchical galactic assembly models, galaxies containing SMBHs merge frequently, and the SMBHs at their centers eventually form a population of SMBHBs~\citep[see e.g.][]{kormendy_inward_1995a,magorrian_demography_1998a,ferrarese_supermassive_2005a,kormendy_coevolution_2013, begelman_massive_1980,volonteri_assembly_2003}. The distribution of SMBHBs over redshift $z$ is therefore expected to trace major galaxy mergers~(e.g.~\citealt{begelman_massive_1980};~\citetalias{sesana_systematic_2013a};~\citealt{simon_constraints_2016}).
        Significant effort has gone into modeling the GWB by deriving the number density of SMBHBs from expected galaxy major merger rates~(e.g.~\citetalias{sesana_systematic_2013a};~\citealt{simon_constraints_2016,middleton_no_2018};~\citealt{chen_constraining_2019}).
        These models are derived from galaxy observations~\citep[e.g.][]{bell_optical_2003,borch_stellar_2006,drory_bimodal_2009,ilbert_galaxy_2010}, and generally rely on assumed pair fractions~\citep[e.g.][]{bundy_greater_2009, deravel_vimos_2009, lopez-sanjuan_dominant_2012}, galaxy merger timescales~\citep[e.g.][]{kitzbichler_calibration_2008,lotz_effect_2010}, and SMBH -- host galaxy scaling relations~\citep[e.g.][]{mcconnell_revisiting_2013,kormendy_coevolution_2013}.
        However while these galactic major merger models are a powerful tool for constraining SMBHB populations, underlying assumptions can vary considerably between models, thus motivating a complementary approach to modeling SMBHB populations which can similarly be tied to observations.
        
        Importantly, major mergers --  especially gas rich ones -- are believed to trigger quasar activity~\citep{sanders_ultraluminous_1988a,volonteri_assembly_2003,granato_physical_2004a,hopkins_cosmological_2008a}, suggesting it might be possible to derive constraints on the SMBHB population from quasar populations.
        Furthermore, there have been numerous observations of dual AGN systems~(e.g.~\citealt{rodriguez_compact_2006a, shen_type_2011a, comerford_origin_2018a, dey_unique_2019a, goulding_discovery_2019},~G19~hereafter; ~\citealt{bhattacharya_detection_2020a, foord_second_2020a, kim_dual_2020a,reines_new_2020, ward_agn_2020, monageng_radio_2021a, severgnini_possible_2021a, shen_hidden_2021a}) in which the SMBHB progenitors are observable as unbound AGN pairs with separations on the order of $\sim 100$~pc, which may also suggest a connection between AGN and SMBHB populations~\citep{goulding_galaxy_2018a}, though the strength of any such connection remains unclear, and is still the matter of much debate \citep{zhao_role_2019,stemo_catalog_2020,silva_galaxy_2021a}.
        The quasar population is generally well-observed and characterized over a large $z$ range~(e.g. \citealt{hopkins_observational_2007, shen_bolometric_2020}, which cover $z = 0 - 6$ and $z = 0 - 7$, respectively), making it an ideal observational anchor for the SMBHB population.
        
        Several AGN-based models of SMBHB populations already exist, however these models are either based on specific subpopulations of AGN~(e.g. \citealt{holgado_pulsar_2018}, which uses blazars; \citetalias{goulding_discovery_2019}, which uses dual AGN), or use quasars to indirectly constrain SMBHB populations via e.g. halo merger trees with seed black holes~\citep{volonteri_assembly_2003,sesana_lowfrequency_2004a,volonteri_subparsec_2009a} or large-volume hydrodynamic simulations~\citep{volonteri_cosmic_2016a,volonteri_black_2020a}.
        There are also SMBHB population models based on the hypothesis that some SMBHBs may be identifiable as periodically variable quasars~\citep[e.g.][]{xin_ultrashortperiod_2021}, though the properties of this population are only expected to be accessible to future surveys such as the Legacy Survey of Space and Time by the Vera C. Rubin Observatory~\citep{haiman_population_2009,xin_ultrashortperiod_2021}.
        In this work we present an exploratory, semi-analytic and self-consistent model of the GWB based on broad, empirically characterized quasar populations~\citep[e.g.][]{hopkins_observational_2007}, where quasar is defined to include all AGN with X-ray luminosities above $\sim 10^{42} \; \rm{erg} \; \rm{s}^{-1}$.
        Given that individual SMBHB systems are expected to be observable via nHz GWs in the near future~(\citealt{rosado_expected_2015a}; \citealt{mingarelli_local_2017a}, M17 hereafter), our model makes explicit use of the local number density of SMBHBs as a parameter, allowing it to be tied to local observations and to self-consistently predict the local number density of SMBHBs, given the anticipated measurement of the GWB amplitude.
        
        This paper is organized as follows:
        in Section~\ref{sec:methods} we present two types of models, reviewing major merger models of SMBHBs in Section~\ref{sec:major_mergers} and developing our quasar-based model in Section~\ref{sec:agn_model}. In Section~\ref{sec:results} we compare the SMBHB populations implied by each type of model, and discuss their astrophysical interpretations.
        Considering the large amplitude of the potential NANOGrav 12.5-yr GWB signal, in Section~\ref{sec:m-m_bulge} we check if using a heavy-set $M_{\rm{BH}} - M_{\rm{bulge}}$ relation with the major merger formalism of \citetalias{mingarelli_local_2017a} could explain such a large signal.
        The main results of this paper are summarized in Section~\ref{sec:conclusion}.

        Throughout this work we use geometric units wherein $G = c = 1$.

    \section{Modeling the GWB}
    \label{sec:methods}
    
        \begin{figure*}
            \centering
            \includegraphics[width=\textwidth]{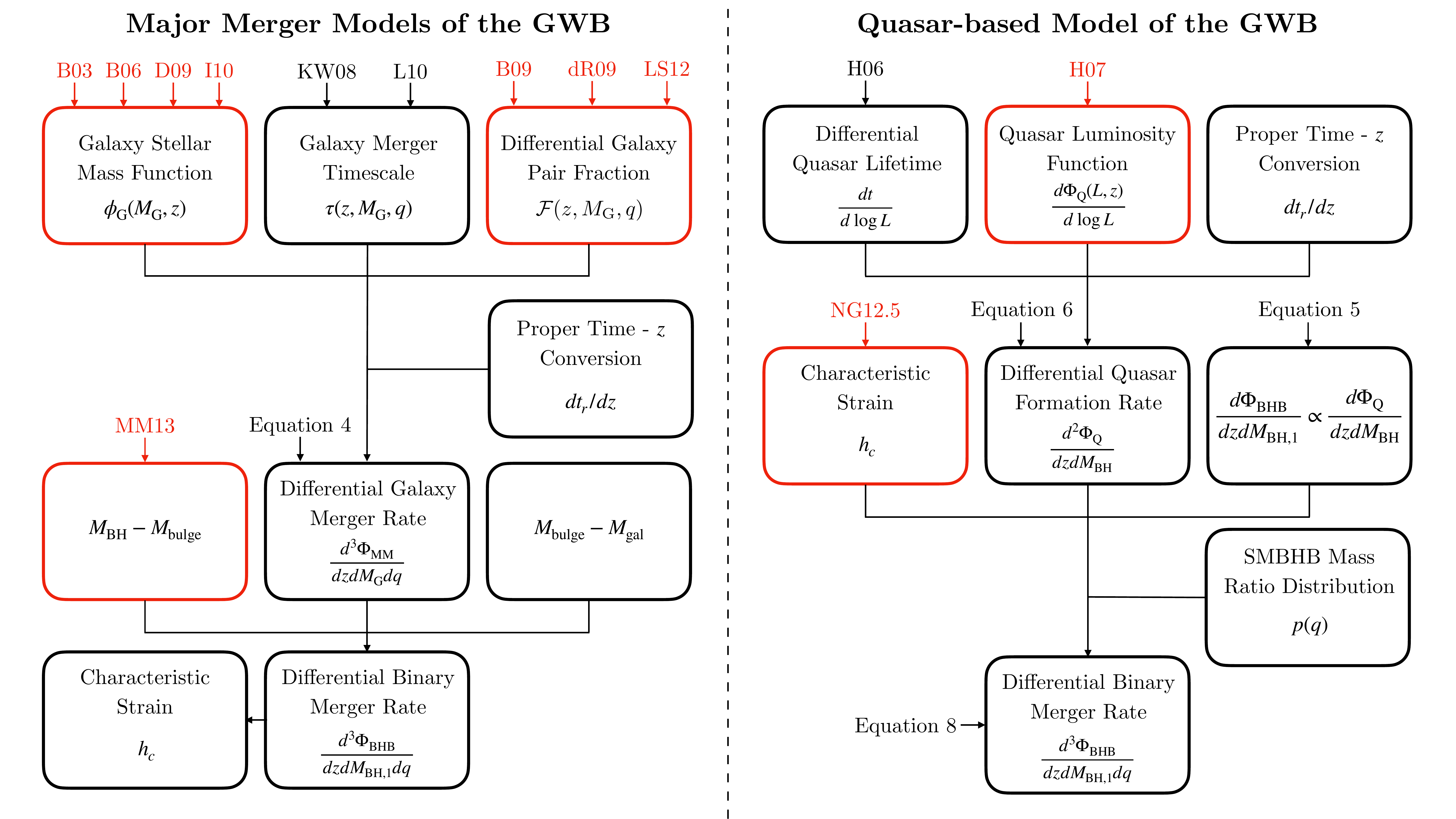}
            \caption{Flowchart comparing the construction of major merger models and our quasar-based model. Black represents theoretical models and assumptions, while observables and empirically constrained models are red, though we note the differential galaxy pair fraction includes theoretical assumptions about the mass ratio distribution. {\em Left:} The major merger models we consider assume galaxy stellar mass functions from \citet[B03]{bell_optical_2003}, \citet[B06]{borch_stellar_2006}, \citet[D09]{drory_bimodal_2009}, and \citet[I10]{ilbert_galaxy_2010}, galaxy merger timescales from \citet[KW08]{kitzbichler_calibration_2008} and \citet[L10]{lotz_effect_2010}, pair fractions from \citet[B09]{bundy_greater_2009}, \citet[dR09]{deravel_vimos_2009}, and \citet[LS12]{lopez-sanjuan_dominant_2012}, and the $M_{\rm{BH}} - M_{\rm{bulge}}$ relation from \citet[MM13]{mcconnell_revisiting_2013}. See \Cref{tab:s13_combos} for the specific model combinations used. {\em Right:} Our quasar-based model assumes the differential quasar lifetime from \citet[H06]{hopkins_unified_2006}, the QLF from \citet[H07]{hopkins_observational_2007}, and the characteristic strain from \citet[NG12.5]{arzoumanian_nanograv_2020a}.}
            \label{fig:model_flowchart}
        \end{figure*}
        
        The cosmic population of SMBHBs can be characterized in terms of the mass of the primary SMBH, $M_{\rm{BH}, 1}$, the mass of the secondary, $M_{\rm{BH}, 2}$, their mass ratios, $q = M_{\rm{BH}, 2} / M_{\rm{BH}, 1} \leq 1$, and $z$.
        The characteristic strain, or characteristic amplitude, of the nHz GWB generated by this population is given by~(e.g. \citealt{phinney_practical_2001a,sesana_stochastic_2008};~\citetalias{sesana_systematic_2013a})
            \begin{multline}
            \label{eq:strain}
                 h_{c}^{2}(f) = \frac{4}{\pi f^{2}} \iiint \dd{M_{\rm{BH}, 1}} \dd{z} \dd{q} \\
                 \times \frac{\dd[3]{\Phi_{\rm{BHB}}}}{\dd{M_{\rm{BH}, 1}} \dd{z} \dd{q}} \frac{1}{1 + z} \dv{E_{\rm{gw}}}{\ln f_{r}},
            \end{multline}
        where $\dv*{E_{\rm{gw}}}{\ln f_{r}}$ is the GW energy emitted per logarithmic frequency interval, $f_{r} = f (1 + z)$ is the frequency of emission in the rest frame of the source, and $\dd[3]{\Phi_{\rm{BHB}}} / \dd{M_{\rm{BH}, 1}} \dd{z} \dd{q}$ is the differential comoving number density of SMBHBs per unit mass, $z$, and mass ratio, i.e. the SMBHB population model.
        For a population of circular SMBHBs, the energy emitted in GWs per logarithmic frequency interval is~\citep{thorne_gravitational_1987a,phinney_practical_2001a}
        \begin{equation}
        \label{eq:energy_differential}
            \dv{E_{\rm{gw}}}{\ln f_{r}} = \frac{\left(\pi f_{r}\right)^{2 / 3}}{3} \mathcal{M}^{5 / 3},
        \end{equation}
        where 
        \begin{equation}
        \label{eq:chirp_mass}
            \mathcal{M}^{5 / 3} = \frac{q}{(1 + q)^{2}} M_{\rm{BHB}}^{5 / 3}
        \end{equation}
        is the SMBHB chirp mass, and $M_{\rm{BHB}} = M_{\rm{BH}, 1} + M_{\rm{BH}, 2}$ is the total binary mass. The differential binary merger rate,
        $\dd[3]{\Phi_{\rm{BHB}}} / \dd{M_{\rm{BH}, 1}} \dd{z} \dd{q}$, is therefore the main empirical constraint on the characteristic strain, with models of this distribution generally being constrained through both theory and observations.
        
        In the following sections we detail the construction of major merger models which we compare to our new quasar-based model, with a summary presented in \Cref{fig:model_flowchart}. To construct SMBHB populations, major merger models assume galaxy stellar mass functions, galaxy merger timescales, galaxy pair fractions, and SMBH-host scaling relations.
        Our quasar-based model instead assumes a quasar luminosity function (QLF), a differential quasar lifetime, and proportionality between the SMBHB and quasar population distribution, which is in turn constrained by the characteristic strain of the GWB. The details are given below.
        
        \subsection{Major Merger Models}
        \label{sec:major_mergers}
        
            \begin{table*}[hbtp!]
            \centering
            \begin{tabular}{c|c|c|c|c}
                Model Name & $\phi_{\rm{G}}(M_{\rm{G}}, z)$ & $f_{p}(M_{\rm{G}}, z)$ & $\tau(z, M_{\rm{G}}, q)$ & $z_{\max}$ \\
                 \hline
                S1 & \citet{borch_stellar_2006} & \citet{bundy_greater_2009} & \citet{kitzbichler_calibration_2008} & 0.9 \\
                S2 & \citet{drory_bimodal_2009}\footnote{\label{note:local}matched locally to \citet{bell_optical_2003}.} & \citet{deravel_vimos_2009} & \citet{lotz_effect_2010} & 0.9 \\
                S3 & \citet{ilbert_galaxy_2010}\footref{note:local} & \citet{lopez-sanjuan_dominant_2012} & \citet{kitzbichler_calibration_2008} & 1.75 \\
            \end{tabular}
            \caption{The three combinations of models from \citetalias{sesana_systematic_2013a} used in this work. For all models, the $M_{\rm{BH}} - M_{\rm{bulge}}$ relationship from \citet{mcconnell_revisiting_2013} is used. While non-exhaustive, these are considered to generally encompass the range in galaxy observations used.}
            \label{tab:s13_combos}
        \end{table*}
        
            Given that SMBHBs form as the result of galaxy major mergers~\citep{begelman_massive_1980}, constraints on their population have frequently been derived from observationally constrained major merger rates which we briefly review here (e.g.~\citetalias{sesana_systematic_2013a};~\citealt{simon_constraints_2016}).
            These models derive the galaxy major merger rate from the observed distribution of galaxies over mass and $z$, combined with empirical pair fractions and theoretical merger timescales, as shown on the left-hand side of \Cref{fig:model_flowchart}.
            This is then converted to a SMBHB merger rate by assuming SMBH-host galaxy scaling relations to convert from galaxy stellar mass to SMBH mass.
            
            This galaxy merger rate (i.e. the differential merger rate density) is modeled as
            \begin{equation}
                \label{eq:diff_galaxy_merger_rate}
                \frac{\dd[3] \Phi_{\rm{MM}}}{\dd{z} \dd{M_{\mathrm{G}}} \dd{q}} = \frac{\phi_{\rm{G}}(M_{\rm{G}}, z)}{M_{\rm{G}} \ln 10} \frac{\mathcal{F}(z, M_{\rm{G}}, q)}{\tau(z, M_{\rm{G}}, q)} \dv{t_{r}}{z},
            \end{equation}
            where $\phi_{\rm{G}}(M_{\rm{G}}, z) = \dv*{\Phi_{\rm{G}}(M_{\rm{G}}, z)}{\log M_{\rm{G}}}$ is the galaxy mass function at $z$, $\mathcal{F}(z, M_{\rm{G}}, q)$ is the differential pair fraction of galaxies per mass ratio, $\tau(z, M_{\rm{G}}, q)$ is the typical galaxy merger timescale, and $\dv*{t_{r}}{z}$ converts from a proper time rate to a $z$ distribution as set by standard cosmology~\citepalias{sesana_systematic_2013a}. 
            The differential pair fraction is taken to be $\mathcal{F}(z, M_{\rm{G}}, q) = \dv*{f(M_{\rm{G}}, z)}{q} = - f_{p}(z) / (q \ln q_{m})$, where $f_{p}(z)$ is an assumed pair fraction and $q_{m}$ a minimum mass ratio.
            The galaxy masses are then converted to SMBHB masses by assuming an SMBH -- host galaxy relation, such as $M_{\rm{BH}} - M_{\rm{bulge}}$. To account for the delay between the start of the galaxy merger and the SMBHs forming a binary, $\phi_{\rm{G}}(M_{\rm{G}}, z)$ and $\mathcal{F}(z, M_{\rm{G}}, q)$ are evaluated at $\phi(M_{\rm{G}}, z + \delta z)$ and $\mathcal{F}(z + \delta z, M_{\rm{G}}, q)$, respectively, where $\delta z$ is a redshift delay corresponding to a galaxy merger timescale $\tau$.
            \citetalias{sesana_systematic_2013a} notes that this model implicitly assumes that all SMBHBs coalesce instantaneously at the merger time of the hosts, though in actuality galaxy merger timescales may differ from SMBHB merger timescales.
            A detailed derivation of \Cref{eq:diff_galaxy_merger_rate} is found in \citetalias{sesana_systematic_2013a}.
            
            To understand the range in SMBHB populations predicted by major mergers, we select three realizations of models, referred to as S1, S2, and S3, from \citetalias{sesana_systematic_2013a} to characterize the broad range of possible SMBHB populations.
            We choose combinations of the galaxy mass functions, pair fractions, and timescales used therein to cover a representative range of major merger model behavior. This is summarized in \Cref{tab:s13_combos}.
            This was not done exhaustively: we only attempt to choose combinations of pair fractions and timescales such that S1 and S2 minimize and maximize, respectively,  the SMBHB merger rate at $z = 0$.  S3 incorporates the pair fraction from \citet{lopez-sanjuan_dominant_2012}. 
            We emphasize that the models chosen here represent only a small subset of the 216 model combinations considered by \citetalias{sesana_systematic_2013a}, and are intended to train our intuition.
            
            For all models, we assume the $M_{\rm{BH}} - M_{\rm{bulge}}$ relation from \citet{mcconnell_revisiting_2013}.
            Up to our choice of $M_{\rm{BH}} - M_{\rm{bulge}}$, we consider the models chosen to generally encompass the range of models considered in \citetalias{sesana_systematic_2013a}, with the exception of the pair fraction from \citet{xu_majormerger_2012}, which uses a local pair fraction observation that is likely unrealistic~\citep{robotham_galaxy_2014}.
            The galaxy mass functions used are all found to be similar over $0 \leq z \lesssim 1$, so as to not significantly affect these choices of combinations.
            We emphasize that the models used here are only a subset of the models \citetalias{sesana_systematic_2013a} considers.
        
        \subsection{New Quasar-Based Model}
        \label{sec:agn_model}
        
        \begin{table*}[t]
            \centering
            \begin{tabular}{c|c|c}
                Symbol & Description & Units \\
                \hline
                $z$ & Redshift & - \\
                $M_{\mathrm{BH}}$ & BH mass & $\mathrm{M}_{\odot}$ \\
                $\mathcal{M}$ & Chirp mass & $\mathrm{M}_{\odot}$ \\
                $\mathcal{M}_{\rm{det}}$ ($M_{\rm{BH,det}}$) & Minimum detectable chirp mass (minimum detectable BH mass) & $\mathrm{M}_{\odot}$ \\
                $q$ & Mass ratio & - \\
                $\frac{\dd[3]{\Phi_{\mathrm{BHB}}}}{\dd{M_{\mathrm{BH}, 1}} \dd{z} \dd{q}}$ & Differential comoving number density of SMBHBs (i.e. SMBHB population model) & $\mathrm{M}_{\odot}^{-1} \rm{Mpc}^{-3}$  \\
                $\Phi_{\mathrm{BHB}, 0}$ & Local comoving number density of SMBHBs  & $\rm{Mpc}^{-3}$ \\
                $\Phi_{\mathrm{Q}, 0}$ & Local comoving number density of quasars  & $\rm{Mpc}^{-3}$ \\
                $f_{\rm{BHB}}(z = 0 | \rm{Quasar})$ & The local fraction of quasars with associated SMBHBs & - \\
                $f_{\rm{Q}}(z = 0 | \rm{BHB})$ & The local fraction of SMBHBs with associated quasars & - \\
                \hline
            \end{tabular}
            \caption{Reference of symbols used throughout this work with descriptions and units.}
            \label{tab:symbols}
        \end{table*}
            
            While galaxy major merger models are a powerful tool for constraining SMBHB populations, they necessarily make assumptions about e.g., the $M_{\rm{BH}} - M_{\rm{bulge}}$ relation~\citep[e.g.][]{mcconnell_revisiting_2013,kormendy_coevolution_2013}, galaxy merger timescales~\citep[e.g.][]{kitzbichler_calibration_2008,lotz_effect_2010}, and galaxy pair fractions~\citep[e.g.][]{bundy_greater_2009,deravel_vimos_2009,lopez-sanjuan_dominant_2012}, which can all vary considerably.
           Another interesting signpost to galaxy mergers -- and therefore SMBH mergers -- are quasars, which in some cases are driven by mergers ~\citep{sanders_ultraluminous_1988a,granato_physical_2004a,hopkins_unified_2006,hopkins_cosmological_2008a}.
            This suggests it might be possible to link SMBHBs to quasar populations.
            Here we explore the construction of such a model, summarized on the right-hand side of \Cref{fig:model_flowchart}, which instead assumes quasars provide an indirect tracer of the SMBH merger rate.
            The black hole mass function (BHMF) is derived from the observed QLF combined with an estimate for the Eddington ratio distribution of the lifetime of the quasar from the galaxy merger simulations in \citet{hopkins_unified_2006}.
            We then assume the ratio of quasars and SMBHBs over cosmic time is fixed, and draw the SMBHB mass ratios from a distribution of merger mass ratios $p(q)$.
            Our model is matched to an assumed GWB amplitude \citep{arzoumanian_nanograv_2020a} to self-consistently predict the SMBHB population.
            
            To construct our quasar-based model, we make the simplifying assumption
            \begin{equation}
            \label{eq:agn_prop}
                \frac{d^{3} \Phi_{\rm{BHB}}}{dM_{\rm{BH}, 1} dz dq} \propto \frac{\dd[2]{\Phi_{\rm{Q}}}}{\dd{M_{\rm{BH}}}\dd{z}} p(q),
            \end{equation}
            i.e. the number density of SMBHBs is proportional to the number density of quasars, $\dd[2]{\Phi_{\rm{Q}}} / \dd{M_{\rm{BH}}}\dd{z}$, and an assumed mass-ratio distribution $p(q)$.
            We take $p(q)$ to be log-normal, with a peak at $\log q = 0$ and $\sigma_{\log q} = 0.5 \; \rm{dex}$, similar to the ``hiq" model in \citealt{sesana_testing_2018}. 
            To calculate the quasar number density, we convert the differential quasar triggering rate, $\dd[2]{\Phi_{\rm{Q}}} / \dd{M_{\rm{BH}}}\dd{t_{r}}$, from \citet{hopkins_observational_2007} to a $z$ distribution by taking
            \begin{equation}
            \label{eq:triggering_rate}
                \frac{\dd[2]{\Phi_{\rm{Q}}}}{\dd{M_{\rm{BH}}}\dd{z}} = \frac{\dd[2]{\Phi_{\rm{Q}}}}{\dd{M_{\rm{BH}}}\dd{t_{r}}} \dv{t_{r}}{z},
            \end{equation}
            where $\dv*{t_{r}}{z}$ is set by standard cosmology.
            This conversion is analogous to the conversion between SMBHB merger rate and number density noted for the general case by \citet{phinney_practical_2001a}, and for galaxy major merger models by \citetalias{sesana_systematic_2013a}.
            For ease of reference a summary of the symbols used throughout this section is provided in \Cref{tab:symbols}.

            Our assumption in \Cref{eq:agn_prop} is similar in spirit to the assumptions made in \citetalias{goulding_discovery_2019}.
            In that work, the number density of SMBHBs similar to the dual AGN J1010+1414 (a SMBHB progenitor which could be emitting nHz GWs today) was derived by normalizing the observed quasar number density (via the QLF) to the predicted number density of dual AGN similar to J1010+1413 at $z \sim 0.2$.
            Likewise, we expect that nHz GWs from individual SMBHBs will be observable in the next 5-10 years~(\citealt{rosado_expected_2015a}, \citetalias{mingarelli_local_2017a}, \citealt{kelley_single_2018}, \citealt{xin_multimessenger_2020}), just as some dual AGN, such as J1010+1413, are observable electromagnetically now.
            Therefore, in the same way that dual AGN observations can be used to constrain their number density, we will also be able to directly constrain the number density of SMBHBs, $\Phi_{\rm{BHB}, 0}$.
            We use the amplitude of the common red-noise process from the NANOGrav 12.5-yr data as an estimate for $h_{c}$, from which we can constrain the local number density of SMBHBs, $\Phi_{\mathrm{BHB}, 0}$, since $h^2_{c} \propto \Phi_{\mathrm{BHB}, 0}$. This allows us to re-normalize the observed quasar distribution, as in \citetalias{goulding_discovery_2019}.
            
            Since we also expect to detect the most massive SMBHBs first, we set a lower detectable chirp mass, $\log(\mathcal{M}_{\rm{det}}  / \rm{M}_{\odot}) \approx 8$, when calculating $\Phi_{\rm{BHB}, 0}$. This is quite conservative, since \citetalias{mingarelli_local_2017a} found that even within 225 Mpc SMBHBs with chirp mass $\lesssim 10^8~M_\odot$ were infrequently detectable.
            
            Concretely, we define the local number density of SMBHBs as
            \begin{equation}
            \label{eq:norm}
                \Phi_{\rm{BHB}, 0} = \iint_{\substack{M_{\rm{BH}, 1} \geq M_{\rm{BH, det}}, \\ q \geq 0.25}} \dd{M_{\rm{BH}}} \dd{q} \eval{\frac{d^{3} \Phi_{\rm{BHB}}}{dM_{\rm{BH}, 1} dz dq}}_{z = 0}.
            \end{equation}
            where $M_{\rm{BH}, 1} \geq M_{\rm{BH, det}}$ is functionally equivalent to $\mathcal{M} \geq \mathcal{M}_{\rm{det}}$, assuming $q \geq 0.25$. Combined with our proportionality assumption in \Cref{eq:agn_prop}, the normalization condition in \Cref{eq:norm} implies that our SMBHB population model can be written down as
            \begin{equation}
            \label{eq:agn_model}
                \frac{\dd[3]{\Phi_{\rm{BHB}}}}{\dd{M_{\rm{BH}, 1}} \dd{z} \dd{q}} = \frac{\Phi_{\rm{BHB}, 0}}{\Phi_{\rm{Q}, 0}} \frac{\dd[2]{\Phi_{\rm{Q}}}}{\dd{M_{\rm{BH}}}\dd{t_{r}}} \dv{t_{r}}{z} \frac{p(q)}{\int_{q > 0.25} \dd{q} p(q)},
            \end{equation}
            where
             \begin{equation}
            \label{eq:local_agn}
                \Phi_{\rm{Q}, 0} = \int_{M_{\rm{BH}} \geq M_{\rm{BH, det}}} \dd{M_{\rm{BH}}} \eval{\frac{\dd[2]{\Phi_{\rm{Q}}}}{\dd{M_{\rm{BH}}}\dd{t_{r}}} \dv{t_{r}}{z}}_{z = 0}
            \end{equation}
            is the local number density of quasars with mass $M_{\rm{BH}} \geq M_{\rm{BH, det}}$, and $\log (M_{\rm{BH, det}} / \rm{M}_{\odot}) \approx 8.5$.
            If we plug \Cref{eq:agn_model} into the right hand side of \Cref{eq:strain} and assume the NANOGrav 12.5-yr signal~\citep{arzoumanian_nanograv_2020a} as a GWB characteristic strain, we can calculate the value of $\Phi_{\rm{BHB}, 0}$ implied by $h_{c}$, resulting in a self-consistent model of the GWB.
           
            At its core, our approach is simply to renormalize the quasar distribution to create a SMBHB distribution which can be matched to a characteristic strain measurement.
            A similar approach was taken in \citet{sesana_testing_2018}, which also assumes proportionality between SMBHB and quasar populations.
            However, in that work the normalization for the mass function was inferred from potentially periodic quasars under the SMBHB hypothesis, with comparisons then drawn to limits on the GWB set by PTAs.
            In this work we instead use a possible GWB signal to infer the local number density of SMBHBs.
            
            In fact, we do not assume a binary quasar fraction \textit{a priori}, though we do assume the ratio of SMBHBs to quasars is fixed.
            This ratio, which is determined by fitting $\Phi_{\rm{BHB}, 0}$ to the NANOGrav 12.5-yr signal, also ties together the local fraction of quasars associated with SMBHBs, $f_{\rm{BHB}}(z = 0 | \rm{Quasar}) = \Phi_{\rm{BHB, Q}, 0} / \Phi_{\rm{Q}, 0}$, and the local fraction of SMBHBs with associated quasars, $f_{\rm{Q}}(z = 0 | \rm{BHB}) = \Phi_{\rm{BHB, Q}, 0} / \Phi_{\rm{BHB}, 0}$, where $\Phi_{\rm{BHB, Q}, 0}$ is the number density of SMBHB systems with quasar activity.
            $\Phi_{\rm{BHB}, 0} / \Phi_{\rm{Q}, 0}$ thus takes the form
            \begin{equation}
            \label{eq:duty_cycle_frac}
                \frac{\Phi_{\rm{BHB}, 0}}{\Phi_{\rm{Q}, 0}} = \frac{f_{\rm{BHB}}(z = 0 | \rm{Quasar})}{f_{\rm{Q}}(z = 0 | \rm{BHB})},
            \end{equation}
            with the details of this derivation given in Appendix~\ref{sec:pop_frac}.
            While this cannot be used to constrain either fraction alone, it can be used to extend constraints on one fraction to the other.
            For example, since $f_{\rm{BHB}}(z = 0 | \rm{Quasar}) \leq 1$ and $f_{\rm{Q}}(z = 0 | \rm{BHB}) \leq 1$, any non-unity value of $\Phi_{\rm{BHB}, 0} / \Phi_{\rm{Q}, 0}$ will directly imply an upper limit on either $f_{\rm{BHB}}(z = 0 | \rm{Quasar})$, if $\Phi_{\rm{BHB}, 0} / \Phi_{\rm{Q}, 0} < 1$, or $f_{\rm{Q}}(z = 0 | \rm{BHB})$, if $\Phi_{\rm{BHB}, 0} / \Phi_{\rm{Q}, 0} > 1$.
            
            For our SMBHB distribution, we assume the quasar triggering rate from \citet[Table 5, ``Variable $\gamma_{2}$" model]{hopkins_observational_2007}.
            This distribution is derived from the empirical QLF presented in the same work, which is a function of bolometric luminosity and $z$.
            They note that the QLF is the convolution of the quasar triggering rate with the differential quasar lifetime, i.e. the period of time that a quasar radiates at some luminosity, which was previously determined via hydrodynamic simulations of merging galaxies for a range of SMBH masses in \citet{hopkins_unified_2006}.
            The quasar triggering rate as a function of mass and $z$ is then the deconvolution of this differential quasar lifetime, which implicitly accounts for the distribution of Eddington ratios, with the QLF, effectively converting between luminosity and mass.
            We then convert this quasar triggering rate to a quasar number density per unit mass and $z$ with \Cref{eq:triggering_rate} ---  effectively a $z$ dependent mass function for active black holes.
            An overall update to the work done in \citet{hopkins_observational_2007} was presented in \citet{shen_bolometric_2020}, however they do not provide a parameterized fit of the quasar triggering rate.
            While a fitting methodology is provided, such work is beyond the scope of this paper.

    \section{Results}
    \label{sec:results}

        A brief summary of our methods is given here before jumping to our main results. To characterize our quasar-based model, in Section~\ref{sec:parameters} we first parameterize the characteristic strain in terms of the local number density of potentially detectable sources, $\Phi_{\rm{BHB}, 0}$, as well as the minimum mass, $M_{\rm{BH}, 1, \min}$, and maximum redshift $z_{\max}$ considered when integrating \Cref{eq:strain}.
        Using this parameterization, we calculate the value of $\Phi_{\rm{BHB}, 0}$ implied by the potential NANOGrav 12.5-yr signal.
        Next, in Section~\ref{sec:population}, we compare this quasar-based model to three major merger model realizations, S1, S2, and S3, selected from \citetalias{sesana_systematic_2013a}, and to the dual AGN-based model from \citetalias{goulding_discovery_2019}.
        Finally, motivated by a larger than expected local number density of binaries found in Section~\ref{sec:parameters}, in Section~\ref{sec:m-m_bulge} we consider the effect of changing the assumed $M_{\rm{BH}} - M_{\rm{bulge}}$ relation on the local number density of detectable sources predicted using the methods of \citetalias{mingarelli_local_2017a}.
        We compute $N = 1024$ Monte Carlo realizations of each of the SMBHB population models considered, as well as the NANOGrav 12.5-yr signal amplitude~\citep{arzoumanian_nanograv_2020a}, to estimate uncertainties.

        \subsection{The Local SMBHB Population}
        \label{sec:parameters}
        
            \begin{figure*}[htbp!]
                \centering
                \includegraphics[width=\textwidth]{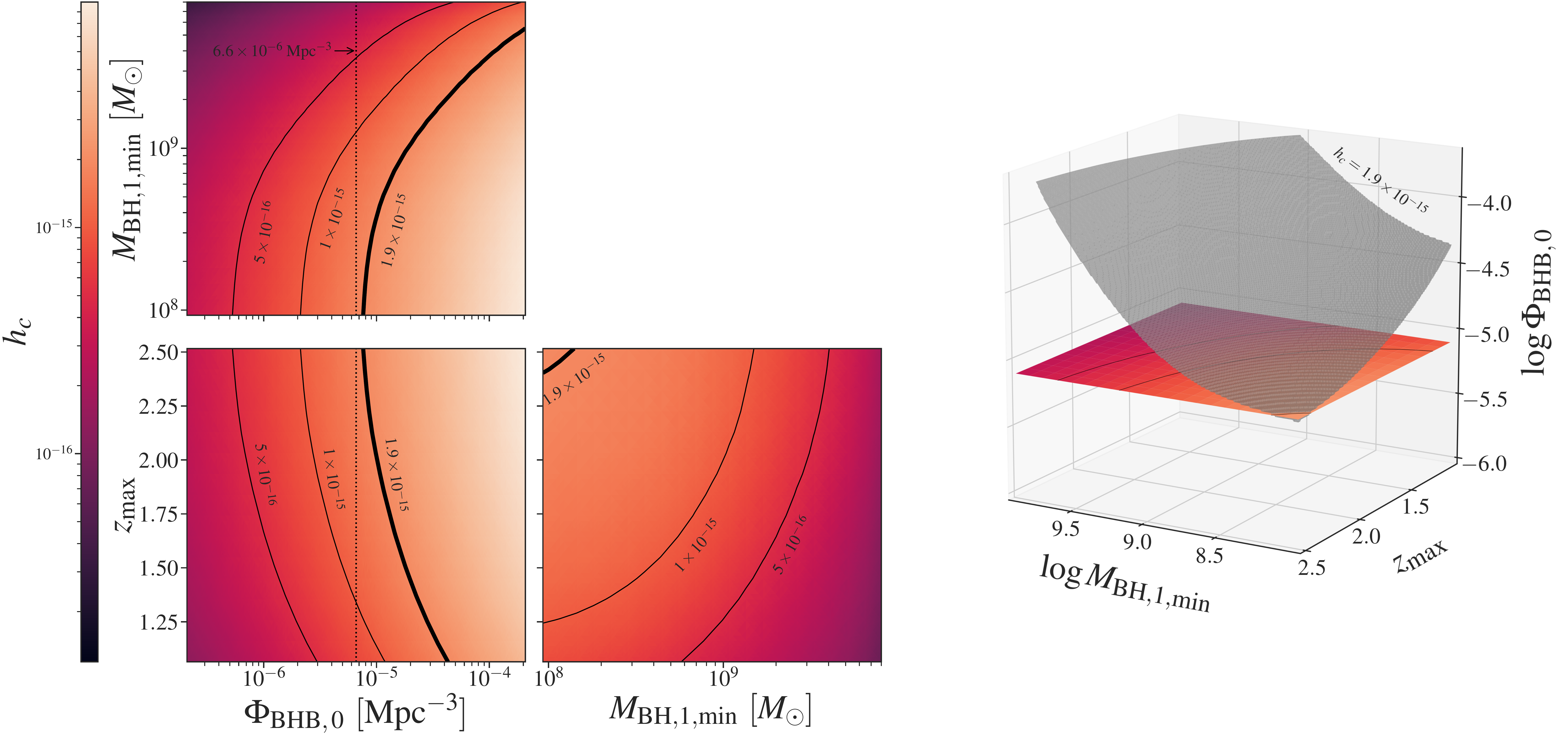}
                \caption{The GWB characteristic strain as a function of the local SMBHB number density, $\Phi_{\rm{BHB}, 0}$, and the minimum mass, $M_{\rm{BH}, 1 \min}$, and $z_{\rm{max}}$, of the population contributing $\gtrsim 95 \%$ of the GWB signal.
                {\em Left:} Three representative slices of the strain in this parameter space, one along each axis.
                Solid contours show their intersection with isosurfaces of constant strain value.
                The NANOGrav 12.5-yr common process signal is shown in bold~\citep{arzoumanian_nanograv_2020a}.
                To recover a similar signal with our quasar-based model, we must have $\Phi_{\rm{BHB}, 0}\approx 6.9 \times 10^{-6}\;\rm{Mpc}^{-3}$, shown as the dotted vertical line in the left panels. 
                Interestingly, this is $\sim5$ times larger than the $1.6 \times 10^{-6}\;\rm{Mpc}^{-3}$ number density predicted using the \citetalias{mingarelli_local_2017a} major merger formalism.
                {\em Right:} 3D visualization of the $z_{\rm{max}} - M_{\rm{BH}, 1, \min}$ panel from the left and its intersection with an $h_{c} = 1.9 \times 10^{-15}$ isosurface in gray.}
                \label{fig:corner}
            \end{figure*}
            
            We estimate the local population of SMBHBs implied by a GWB signal -- such as the common-process signal in the NANOGrav 12.5-yr dataset -- by parameterizing the characteristic strain in terms of the local number density of potentially detectable sources as well as the lower mass and upper $z$ bounds in \Cref{eq:strain}, i.e. $h_{c} = h_{c}(\Phi_{\rm{BHB}, 0}, M_{\rm{BH}, 1, \min}, z_{\max})$, shown in \Cref{fig:corner}.
            This parameterization lets us directly vary the size of the local SMBHB population, noting that in doing so we increase the population of SMBHBs everywhere, based on our model implementation outlined in Section~\ref{sec:agn_model}.
            By varying the bounds of integration as well, we can ensure that for any given population of SMBHBs we are considering at least $95\%$ of the possible signal, though we find little variation in where this saturation occurs over the range of $\Phi_{\rm{BHB}, 0}$ considered. In fact, it may be sufficient to consider only $z_{\max} = 2.5$ and $\log\left(M_{\rm{BH}, 1, \min} / \rm{M}_{\odot} \right) \approx 8$ for all values of $\Phi_{\rm{BHB}, 0}$.
            
            To fully encompass the signal from SMBHBs, we consider the SMBHB population with $z \lesssim 3$ and $\log (M_{\rm{BH}, 1} / \rm{M}_{\odot}) \gtrsim 7$.
            With $h_{c}$ as the GWB amplitude at frequency $f = 1\;\rm{yr}^{-1}$, we find that an amplitude $h_{c} = \left(1.9^{+0.4}_{-0.4}\right)\times 10^{-15}$ signal, with upper and lower bounds indicating the $68\%$ confidence interval, implies $\Phi_{\rm{BHB}, 0} \approx \left(6.6^{+8.3}_{-3.9}\right) \times 10^{-6}\;\rm{Mpc}^{-3}$ for our quasar-based model.
            This also implies roughly $4 \times$ more SMBHBs than quasars locally, and, following \Cref{eq:duty_cycle_frac} and the constraint $f_{\rm{BHB}}(z=0|\rm{Quasar}) \leq 1$, that at most a quarter of SMBHBs have associated quasar activity, i.e. $f_{\rm{Q}}(z=0|\rm{BHB}) \lesssim 0.25$.
            Given the simplistic nature of our model, we consider this constraint to be weak and a more detailed analysis of this relationship is planned for followup study.
            
            For comparison, the $91 \pm 7$ local binaries predicted by \citetalias{mingarelli_local_2017a} imply a local number density $\Phi_{\rm{BHB}, 0} \approx \left(1.6^{+0.1}_{-0.1}\right) \times 10^{-6}\;\rm{Mpc}^{-3}$, roughly five times smaller than the local number density implied by the NANOGrav 12.5-yr signal.
            If the local SMBHB population is much larger than previously predicted, individual SMBHBs may be detected sooner than previously thought. We will follow up this possibility in a future work.
            The local SMBHB estimates made by \citetalias{mingarelli_local_2017a} are quite conservative; therefore a large amplitude GWB might support less conservative assumptions in major merger models.
            A higher than expected SMBHB local number density may also indicate a significant population of eccentric binaries, which have been found to generally increase the local number density of SMBHBs compared to predominantly circularized populations~(Mohite \& Mingarelli in prep).

        \subsection{Comparing SMBHB Populations}
        \label{sec:population}
        
            \begin{figure*}[htbp!]
                \centering
                \includegraphics[width=\textwidth]{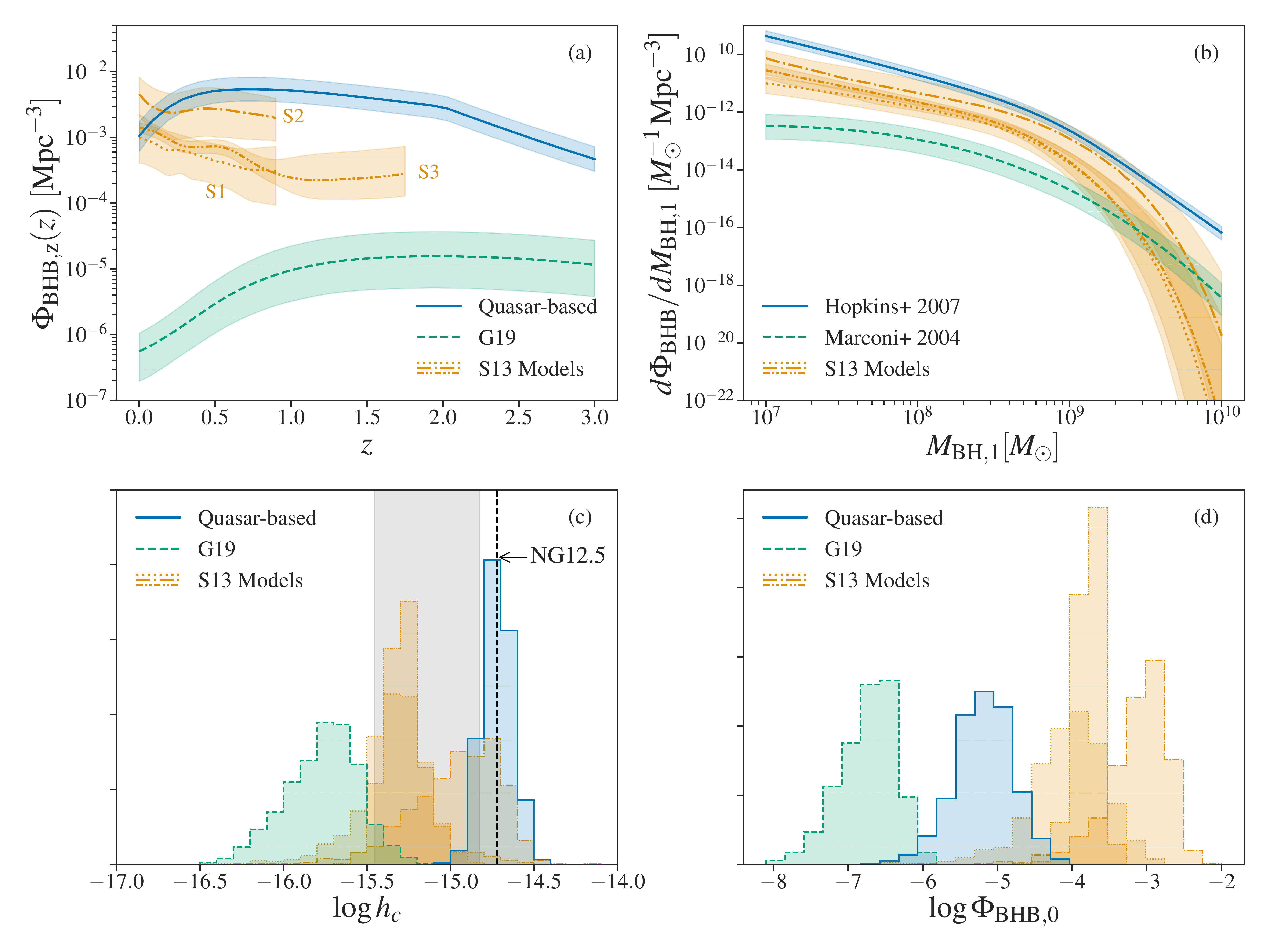}
                \caption{Monte Carlo realizations of SMBHB populations for our quasar-based model (blue solid line) compared to \citetalias{goulding_discovery_2019} (green dashed line) and $3$ model realizations from \citetalias{sesana_systematic_2013a} (orange dotted, dash-dotted, and dash-dot-dotted lines). The galactic mass functions, pair fractions, and timescale models used for the \citetalias{sesana_systematic_2013a} realizations are described in the text.  In panels (a) and (b),  lines denote the medians of each model, while shaded regions denote $68 \%$ confidence intervals. {\em Top left:} Differential number densities of SMBHB populations from each model, marginalized over mass and mass ratio. {\em Top right:} Total BHMFs for each model, i.e. the mass distribution of SMBHBs contributing to the GWB marginalized over mass ratio and $z$. We use the BHMF from \citet{hopkins_observational_2007}, while \citetalias{goulding_discovery_2019} uses \citet{marconi_local_2004}. The \citetalias{sesana_systematic_2013a} BHMFs are derived by combining galactic mass functions with the \citet{mcconnell_revisiting_2013} $M_{\rm{BH}} - M_{\rm{bulge}}$ relation, as detailed in Section~\ref{sec:major_mergers}. {\em Bottom left:} The distribution of characteristic amplitudes for each model, with the $68 \%$ confidence interval from \citetalias{sesana_systematic_2013a} as the gray shaded region ($3.5 \times 10^{-16} < h_{c} < 1.5 \times 10^{-15}$), and the amplitude of the NANOGrav 12.5-yr dataset signal~(NG12.5, \citealt{arzoumanian_nanograv_2020a}) as the vertical dashed line. Our quasar-based model takes a GWB amplitude as input, reproducing the NANOGrav 12.5-yr signal by design. {\em Bottom right:} Local number density distributions.  Our quasar-based model predicts fewer SMBHBs locally than the \citetalias{sesana_systematic_2013a} realizations, despite being matched to a larger GWB than they predict. This is because the \citetalias{sesana_systematic_2013a} realizations' SMBHB distributions peak locally, near $z \sim 0$, while our quasar-based model peaks near $z \sim 0.5$.}
                \label{fig:smbhb_pops}
            \end{figure*}
        
            In \Cref{fig:smbhb_pops} we compare our quasar-based SMBHB population to three major merger model realizations selected from \citetalias{sesana_systematic_2013a} and the dual AGN-based model from \citetalias{goulding_discovery_2019}.
            We emphasize that only our quasar-based model takes the GWB amplitude as input.
            The other models shown are based on electromagnetic observations and theoretical simulations only, without GW input.
            All models include contributions from systems with $\log(M_{\rm{BH}, 1} / \rm{M}_{\odot}) \gtrsim 7$, ensuring that we consider the maximum possible signal,  and $0.25 \leq q \leq 1$.
            For the AGN-based models we choose an upper volume limit $z_{\max} = 3$, while the major merger models are volume limited by the galactic mass functions used~(\citealt{borch_stellar_2006,drory_bimodal_2009,ilbert_galaxy_2010}; see \Cref{tab:s13_combos} for details of $z_{\max}$).
            
            In panel (a) of \Cref{fig:smbhb_pops} we show the number density evolution of sources over $z$, $\Phi_{\rm{BHB}, z}(z) = \eval{\dv*{\Phi_{\rm{BHB}}}{z}}_{z}$, defined analogously to $\Phi_{\rm{BHB}, 0}$ in \Cref{eq:norm}.
            We find that while our quasar-based model and the \citetalias{sesana_systematic_2013a} realizations agree locally, the $z$ evolution of these models are distinct, with our quasar-based model generally predicting larger SMBHB number densities at higher $z$.
            We emphasize, however, the exploratory nature of our model, which likely oversimplifies the relationship between SMBHBs and quasars by assuming only a direct proportionality between their mass and $z$ distributions.
            We expect that explicitly modeling the physical processes linking these populations, instead of normalizing these processes away, will change the details of their distribution.
            The significantly lower number density over $z$ exhibited by the \citetalias{goulding_discovery_2019} dual AGN-based model is expected, given that this model represents a conservative estimate of contributions due to SMBHBs which went through a dual AGN phase similar to J1010+1413.
            
            In panel (b) of \Cref{fig:smbhb_pops} we show the total SMBHB mass function of all models, i.e. the mass distribution of SMBHBs contributing to the GWB integrated over $z$ and $q$, which we consider to broadly agree.
            We emphasize that these distributions represent the overall mass distribution of GWB contributing SMBHBs, and that the distribution of observable SMBHBs will likely differ from these distributions due to mass and $z$ related selection effects (e.g. closer, more massive systems will be easier to detect than less massive systems at higher $z$).
            In panel (c) we show the characteristic strain values associated with each model, noting that the characteristic strain associated with our quasar-based model replicates the NANOGrav 12.5-yr signal {\em by design}.
            The other models shown generally predict characteristic amplitudes below the amplitude of the NANOGrav 12.5-yr signal, with the exception of the model S2, whose $68\%$ confidence interval overlaps significantly with the $68\%$ confidence interval of the NANOGrav 12.5-yr result.
            We can see in panel (a) that, compared to the other \citetalias{sesana_systematic_2013a} realizations, S2 predicts generally higher SMBHB number densities at all $z$, again implying that the NANOGrav 12.5-yr signal might be associated with a SMBHB population that is larger than has generally been predicted.
            
            Finally in panel (d) we show the distribution of $\Phi_{\rm{BHB}, 0}$ values for all models, noting that our quasar-based model predicts a lower local SMBHB number density than the \citetalias{sesana_systematic_2013a} do, despite being matched to a larger GWB amplitude than predicted by the \citetalias{sesana_systematic_2013a} realizations.  We can see in panel (a) that this is likely because our quasar-based model predicts more sources near $z \sim 0.5$ than it does locally,, with correspondingly more contributions to the GWB from these more distant sources.
              The \citetalias{sesana_systematic_2013a} realizations instead all predict the largest concentration of SMBHBs at $z = 0$, leading to larger local SMBHB number densities, and more local contributions to the GWB.
            The values of $h_{c}$ and $\Phi_{\rm{BHB}, 0}$ for all models are given in \Cref{tab:results}.
            
            \begin{figure*}
                \centering
                \includegraphics[width=\textwidth]{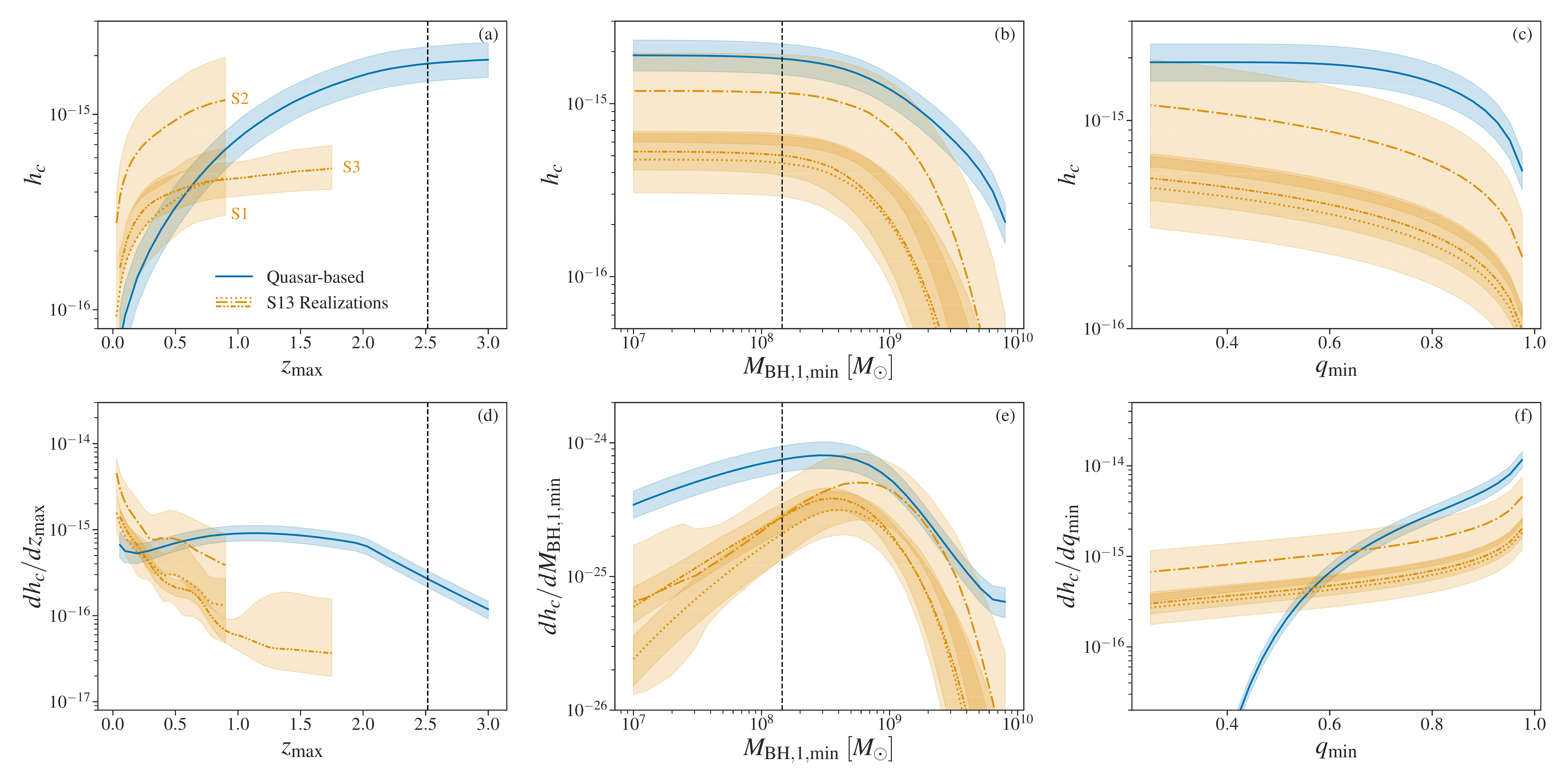}
                \caption{Characteristic strain as a function of integration limits (top row, a-c) and corresponding integrands (bottom row, d-f).  Lines denote the medians of the Monte Carlo realizations, while shaded regions denote $68 \%$ confidence intervals. {\em Left:} Characteristic strain as a function of $z_{\rm{max}}$ (a) and corresponding integrand (d). The notable differences between our quasar-based model and the major merger models is considered to likely be the result of implicitly assuming constant proportionality between the number densities of SMBHBs and quasars. $\gtrsim 95 \%$ of the signal generated by our quasar-based model comes from systems in $z \lesssim 2.5$, shown with vertical dashed lines. {\em Middle:} Characteristic strain as a function of minimum mass (b), and the distribution of GWB contributions by mass (e). For our quasar-based model $\gtrsim 95 \%$ of the signal is generated by systems with $\log(M_{\rm{BH}, 1} / \rm{M}_{\odot}) \gtrsim 8$, consistent with major merger models. {\em Right:} Characteristic strain as a function of minimum mass ratio $q_{\rm{min}}$ (c), and the distribution of contributions (f). We consider the differences between each type of model to largely be due to differences in assumed mass-ratio distributions.}
                \label{fig:systematics}
            \end{figure*}
            
            \begin{table}[htbp!]
            \centering
            \begin{tabular}{c|c|c}
                Model & $h_{c}(f =\SI{1}{\per\year})$ & $\Phi_{\rm{BHB}, 0}\;[\rm{Mpc}^{-1}]$\\
                \hline
                 Quasar-based & $\left( 1.9^{+0.4}_{-0.4}\right) \times 10^{-15}$ & $\left( 6.6^{+8.3}_{-3.9} \right) \times 10^{-6}$ \\
                 G19 & $\left( 1.8^{+0.9}_{-0.7}\right) \times 10^{-16}$ & $\left( 1.1^{+0.9}_{-0.7} \right) \times 10^{-7}$ \\
                 S1 & $\left( 4.7^{+2.1}_{-1.7}\right) \times 10^{-16}$ & $\left( 9.5^{+18.7}_{-6.8} \right) \times 10^{-5}$ \\
                 S2 & $\left( 1.1^{+0.8}_{-0.5}\right) \times 10^{-15}$ & $\left( 8.4^{+9.1}_{-5.7} \right) \times 10^{-4}$ \\
                 S3 & $\left( 5.2^{+1.7}_{-1.1}\right) \times 10^{-16}$ & $\left( 1.7^{+0.6}_{-0.5} \right) \times 10^{-4}$ \\
            \end{tabular}
            \caption{Summary of characteristic strains and local number densities for each evaluated model. Our quasar-based model takes the GWB amplitude as input, reproducing this amplitude by design. Our quasar-based model predicts a larger population of SMBHBs near $z \sim 0.5$ than at $z = 0$, leading to a lower local number density of SMBHBs than predicted by major merger models, which all have population maximums at $z = 0$.}
            \label{tab:results}
        \end{table}
        
            In the top row of \Cref{fig:systematics} we show how the characteristic amplitude of our quasar-based model and the \citetalias{sesana_systematic_2013a} models vary as a function of integration bounds, including $z_{\max}$, $M_{\rm{BH}, 1, \min}$, and $q_{\min}$, while the bottom row of \Cref{fig:systematics} shows the differential contributions to the background along each of these distribution axes.
            
            We find $\gtrsim 95\%$ of the GWB signal generated by our quasar-based model comes from SMBHBs with $\log\left(M_{\rm{BH}, 1} / \rm{M}_{\odot}\right) \gtrsim 8$, shown in panels b and e, and $q_{\min} \gtrsim 0.25$, shown in panels c and f, consistent with the predictions of major merger models~(e.g. \citetalias{sesana_systematic_2013a}; \citealt{simon_constraints_2016}).
            In fact, for the log-normal distribution of $q$ our assumed $q_{\min} = 0.25$ appears to be a generous lower bound, though what constitutes a sufficient lower bound seems to depend on the specific distribution of mass ratios assumed.
            This can be seen in panels c and f as differences between our assumed log-normal distribution and the log-uniform distribution assumed by \citetalias{sesana_systematic_2013a}. 
        
            Examining the volume of the GWB generated by our quasar-based model, we find $\gtrsim 95\%$ of the signal comes from $z_{\max} \sim 2.5$, shown by the vertical dashed line in the left panels of \Cref{fig:systematics}, roughly double the volume predicted by major merger models~(e.g. \citetalias{sesana_systematic_2013a}; \citealt{simon_constraints_2016}).
            We note in panel (a) of \Cref{fig:systematics} that neither the S1 nor S2 models, which are defined to $z \sim 0.9$, appear to fully level off, suggesting that these models may not sufficiently sample the entire volume of the GWB; however it is not clear that sampling to higher $z$ would resolve the volume tension between our quasar-based model and these major merger models.
            In panel (d) of \Cref{fig:systematics} we can see that our quasar-based model generally includes larger contributions from SMBHBs at higher $z$, however we consider the degree of difference to be symptomatic of the simplified relationship between SMBHBs and quasars we have assumed.
        
        \subsection{$M_{\rm{BH}} - M_{\rm{bulge}}$ Relations}
        \label{sec:m-m_bulge}
        
            Here we briefly consider the effect of changing our assumed $M_{\rm{BH}} - M_{\rm{bulge}}$ relation.
            While some of the less conservative predictions of major merger models can generate a GWB similar to the NANOGrav 12.5-yr signal, the majority of these models predict a signal below $1.9 \times 10^{-15}$.
            However, using the \citetalias{sesana_systematic_2013a} formalism, \citet{simon_constraints_2016} found that assuming an $M_{\rm{BH}} - M_{\rm{bulge}}$ model like \citet{kormendy_coevolution_2013}, rather than \citet{mcconnell_revisiting_2013}, resulted in a larger amplitude GWB due to a generally more massive population of SMBHBs.
            
            To test the consistency of this with the major merger formalism of \citetalias{mingarelli_local_2017a}, and to see if a large amplitude GWB might support an $M_{\rm{BH}} - M_{\rm{bulge}}$ relation weighted towards higher masses, we re-run the analysis performed in \citetalias{mingarelli_local_2017a} $100,000$ times using the 2MASS sample from that same work.
            We find $17 \pm 6$ nearby SMBHBs with $\log(M_{BH, 1} / \rm{M}_{\odot}) \gtrsim 8.7$, compared to the $91 \pm 7$ found by \citetalias{mingarelli_local_2017a}.
            Considering that the more massive SMBHs predicted by \citet{kormendy_coevolution_2013} would be stronger GW emitters, we might expect the number of detectable binaries to increase.
            However because the time to coalescence $t_{c}$ is proportional to $\mathcal{M}^{- 5 / 3}$, we find that the number of detectable nearby binaries instead decreases, as the more massive SMBHB systems spend less time in the PTA band. Therefore a given snapshot of the local universe will contain fewer binaries.
            
            The time to detection of individual binaries remains the same, however, as the remaining binaries are still generally more massive, and therefore produce a larger strain. We also note that sources which are not individually detectable are still be expected to contribute to the GWB, which includes GWs from non-resolvable sources.
            
            \begin{figure}[htbp!]
                \centering
                \includegraphics[width=\columnwidth]{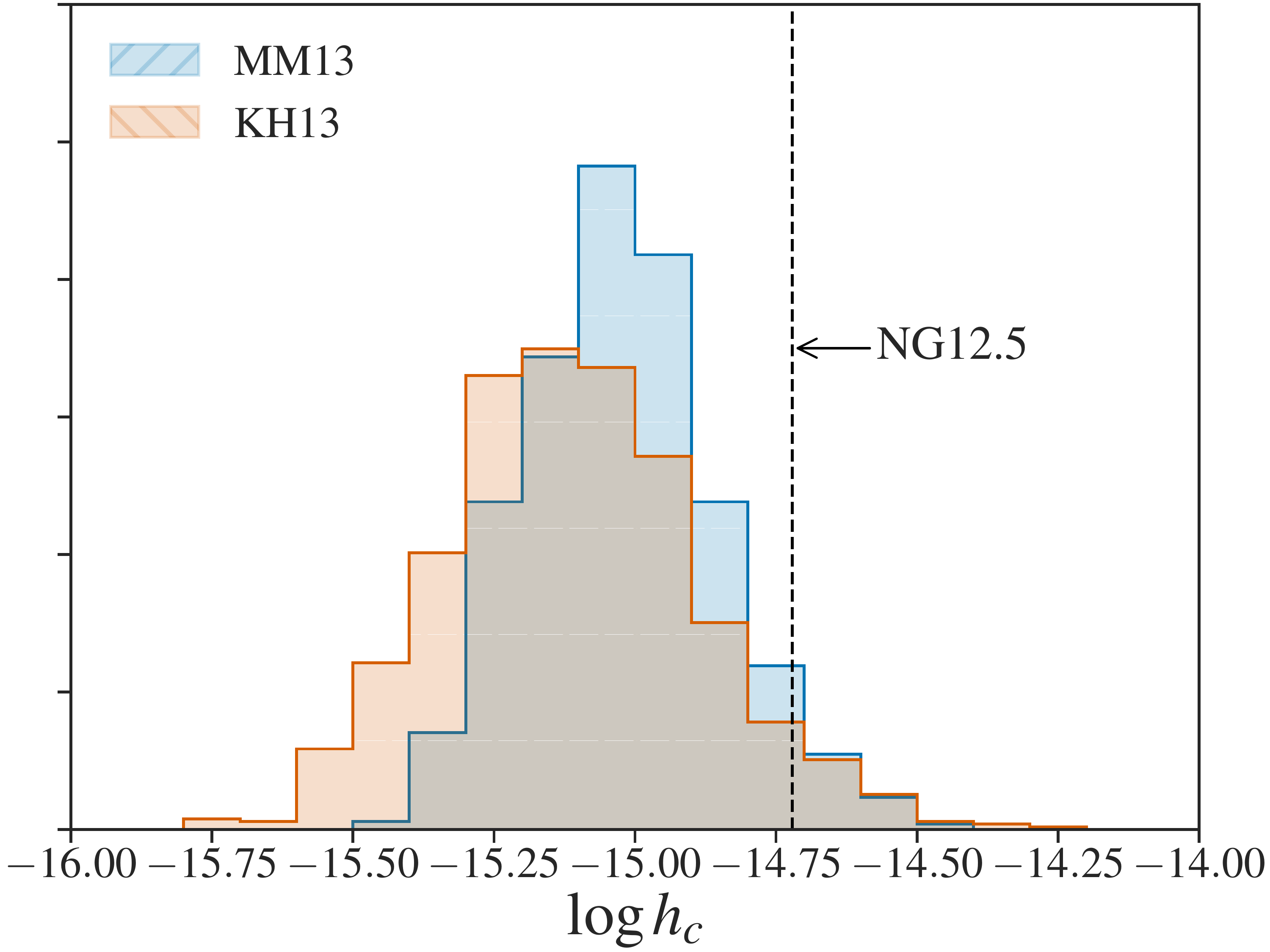}
                \caption{Comparison of quasar-based characteristic strain model realizations under the assumption of \citet{mcconnell_revisiting_2013} (blue) and \citet{kormendy_coevolution_2013} (red) $M_{\rm{BH}} - M_{\rm{bulge}}$ relations. Overall the change in the predicted strain is found to be negligible, with the lower local number density of sources predicted under \citet{kormendy_coevolution_2013} assumptions leading to a slight decrease in the GWB amplitude, despite predicting generally higher mass systems than \citet{mcconnell_revisiting_2013}. Overall both $M_{\rm{BH}} - M_{\rm{bulge}}$ relations predict characteristic strain values that are well within each other's $1 \sigma$ confidence intervals.}
                \label{fig:kormendy_ho}
            \end{figure}
            
            To train one's intuition of how a decrease in the local SMBHB population might change expectations for $h_{c}$, we use these predicted $17 \pm 6$ local SMBHBs to estimate $\Phi_{\rm{BHB}, 0} \approx (3.0^{+1.1}_{-1.1}) \times 10^{-7} \; \rm{Mpc}^{-3}$ in \Cref{eq:agn_model}, sacrificing some self-consistency in the process.
            Estimating $\Phi_{\rm{BHB}, 0} \approx (1.6^{+0.1}_{-0.1}) \times 10^{-6} \; \rm{Mpc}^{-3}$ from \citetalias{mingarelli_local_2017a}, we calculate the characteristic strain associated with each, shown in \Cref{fig:kormendy_ho}.
            As expected, decreasing $\Phi_{\rm{BHB}, 0}$ leads to a decrease in $h_{c}$, though because our quasar-based model does not assume an $M_{\rm{BH}} - M_{\rm{bulge}}$ relation, our result cannot account for any shift in the distribution of SMBHB masses.
            Compared to the increase in $h_{c}$ found by \citet{simon_constraints_2016}, the decrease we find is small, maybe indicating that the increases to $h_{c}$ implied by more massive SMBHBs under \citet{kormendy_coevolution_2013} assumptions are tempered by an overall reduction in the size of the SMBHB population.
            
            We note that differences in how SMBHB merger timescales are computed in each method could also play a role here -- this is also the subject of future work.
            
    \section{Summary}
    \label{sec:conclusion}
    
        We have developed a quasar-based model of the GWB which can be used to self-consistently predict the local number density of SMBHBs, given a measurement of the GWB's characteristic strain \citep[e.g.][]{arzoumanian_nanograv_2020a}.
        Whereas major merger models assume SMBH - host galaxy relations (e.g. $M_{\rm{BH}} - M_{\rm{bulge}}$), galaxy merger timescales, and galaxy pair fractions~\citepalias{sesana_systematic_2013a}, our model instead assumes proportionality between SMBHB and quasar populations, backed by an empirical QLF~\citep{hopkins_observational_2007} and a theoretical differential quasar lifetime~\citep{hopkins_unified_2006}.
        While we plan to apply more sophisticated assumptions in the future, the model presented here offers a view of SMBHB populations complementary to the view offered by galaxy major merger models.
        
        Both our quasar-based model and the S2 major merger model considered from \citetalias{sesana_systematic_2013a} suggest that a SMBHB population giving rise to a GWB similar to the common-process signal in the NANOGrav 12.5-yr dataset may be larger, i.e. more numerous, than has generally been predicted. However the $z$ distribution of our quasar-based model is simplistic in its assumptions, and we thus emphasize its exploratory nature.
        Similar conclusions could likely be drawn from an analysis of the PPTA second data release, which contains a $(2.2^{+0.4}_{-0.3}) \times 10^{-15}$ amplitude signal~\citep{goncharov_evidence_2021a}, implying a slightly larger value of $\Phi_{\rm{BHB}, 0}$ than the NANOGrav 12.5-yr signal. However the overall change to the local number density would be of order unity.
        A generally higher number of SMBHB sources is consistent with findings from \citet{middleton_massive_2021}, which fit the major merger model employed by \citet{chen_constraining_2019} to the NANOGrav 12.5-yr signal. \citet{middleton_massive_2021} found that a GWB like the NANOGrav 12.5-yr common process signal implies a larger SMBHB merger rate normalization than electromagnetic observations alone, consistent with a more numerous SMBHB population. A direct comparison between our model and the \citet{middleton_massive_2021} results will be the subject of future work.
        
        We note it is possible that two of the major merger models investigated here are missing contributions from higher $z$.
        This can be seen for models S1 and S2 in panel (a) of \Cref{fig:systematics}, whose volume-dependent GWB contributions do not appear to level off.
        We also note that contributions from more exotic sources of nHz GWs, such as primordial black holes~\citep[e.g.][]{hawking_gravitationally_1971, carr_black_1974, carr_primordial_1975, vaskonen_did_2021, deluca_nanograv_2020, zhou_primordial_2020a, gao_primordial_2021a, kohri_solarmass_2021} and cosmological phase transitions~\citep[e.g.][]{cline_baryogenesis_2021a, brandenburg_can_2021a, li_nanograv_2021b, kumar_nongaussian_2021}, are not considered in the present work, and thus these sources may also contribute to the GWB.

        We additionally find evidence that an $M_{\rm{BH}} - M_{\rm{bulge}}$ relation weighted towards more massive black holes~\citep[e.g.][]{kormendy_coevolution_2013} under the major merger formalism of \citetalias{mingarelli_local_2017a} may slightly decrease the number density of PTA-band SMBHBs, and by extension the amplitude of the GWB, due to a faster time to coalescence for these systems.
        This is in contrast to results from \citet{simon_constraints_2016}, who found that the larger SMBHBs predicted by the \citet{kormendy_coevolution_2013} $M_{\rm{BH}} - M_{\rm{bulge}}$ relation increased the amplitude of the GWB using the \citetalias{sesana_systematic_2013a} major merger formalism.
        Our result may indicate that any increases to the GWB amplitude due to mass are be tempered by an overall reduction in the number of PTA band SMBHB sources.
        This implies that a top-heavy $M_{\rm{BH}} - M_{\rm{bulge}}$ relation alone~\cite[such as in][]{kormendy_coevolution_2013} may not be enough to account for a large GWB amplitude.
         Furthermore, while the \citet{kormendy_coevolution_2013} scaling relations result in fewer local SMBHBs, the sources predicted probe a higher mass range than under \citet{mcconnell_revisiting_2013} assumptions. As a result, if we also consider less massive SMBHBs, which are more numerous, we find the overall number density of SMBHBs contributing to the background to be similar, regardless of the $M_{\rm{BH}} - M_{\rm{bulge}}$ relation used. In other words, predicting fewer high mass SMBHBs does not necessarily imply fewer less massive SMBHBs.

        While our exploratory quasar-based model presented here may oversimplify the relationship between SMBHBs and quasars, we consider it to be a self-consistent basis for modeling their relationship, and one which can additionally be tied to the expected locally observable population of SMBHBs.
        To construct a more realistic model of SMBHB populations from quasar populations we will likely need to model the quasar pair fraction, which could additionally be constrained with dual AGN observations.
        Upcoming measurements of the GWB by PTA experiments will be crucial for constraining the SMBHB populations of both quasar-based and major merger models, including the local number density of SMBHB sources and their distribution throughout the universe.
    
        \section*{Acknowledgements}
        
        We thank the anonymous referee for useful comments improving this paper. JACC and CMFM thank Luke Kelley, Jeff Hazboun, Siyuan Chen, Nihan Pol, Alberto Sesana, Joe Lazio, the International Pulsar Timing Array Gravitational-Wave Analysis Group and the NANOGrav Astrophysics Working Group for useful comments and discussions. CMFM thanks David A. Forsh and his team for their excellent care. JACC was supported in part by NASA CT Space Grant PTE Federal Award Number 80NSSC20M0129. CMFM and JACC are also supported by the National Science Foundation's NANOGrav Physics Frontier Center, Award Number 2020265. MN was supported by the summer internship program at the Center for Computational Astrophysics at the Flatiron Institute. The Flatiron Institute is supported by the Simons Foundation. Part of this work was done at Jet Propulsion Laboratory, California Institute of Technology, under a contract with the National Aeronautics and Space Administration.
        
        {\em Software:} {\tt J1010\_1413\_gws}\footnote{\href{https://github.com/kpardo/J1010_1413_gws}{github.com/kpardo/J1010\_1413\_gws}}, {\tt NanohertzGWs}~\citep{mingarelli_chiaramingarelli_2017}, {\tt numpy}~\citep{harris2020array}, {\tt scipy}~\citep{2020SciPy-NMeth}, {\tt astropy}~\citep{astropycollaboration_astropy_2013, collaboration_astropy_2018}, {\tt scikit-image}~\citep{scikit-image}.
        
        {\em Codes and Data:} The codes and data used to produce the plots and results in this work, as well as the relevant final data products, are freely available as a collection of Jupyter Notebooks and associated {\tt python} scripts at \href{https://github.com/jacaseyclyde/multimessenger_smbhbs}{github.com/jacaseyclyde/multimessenger\_smbhbs}.

    \bibliographystyle{aasjournal}
    \bibliography{bib}
    
    \appendix
    \section{Population Fraction Derivation}
    \label{sec:pop_frac}
    
    \begin{figure}
        \centering
        \includegraphics[width=\columnwidth]{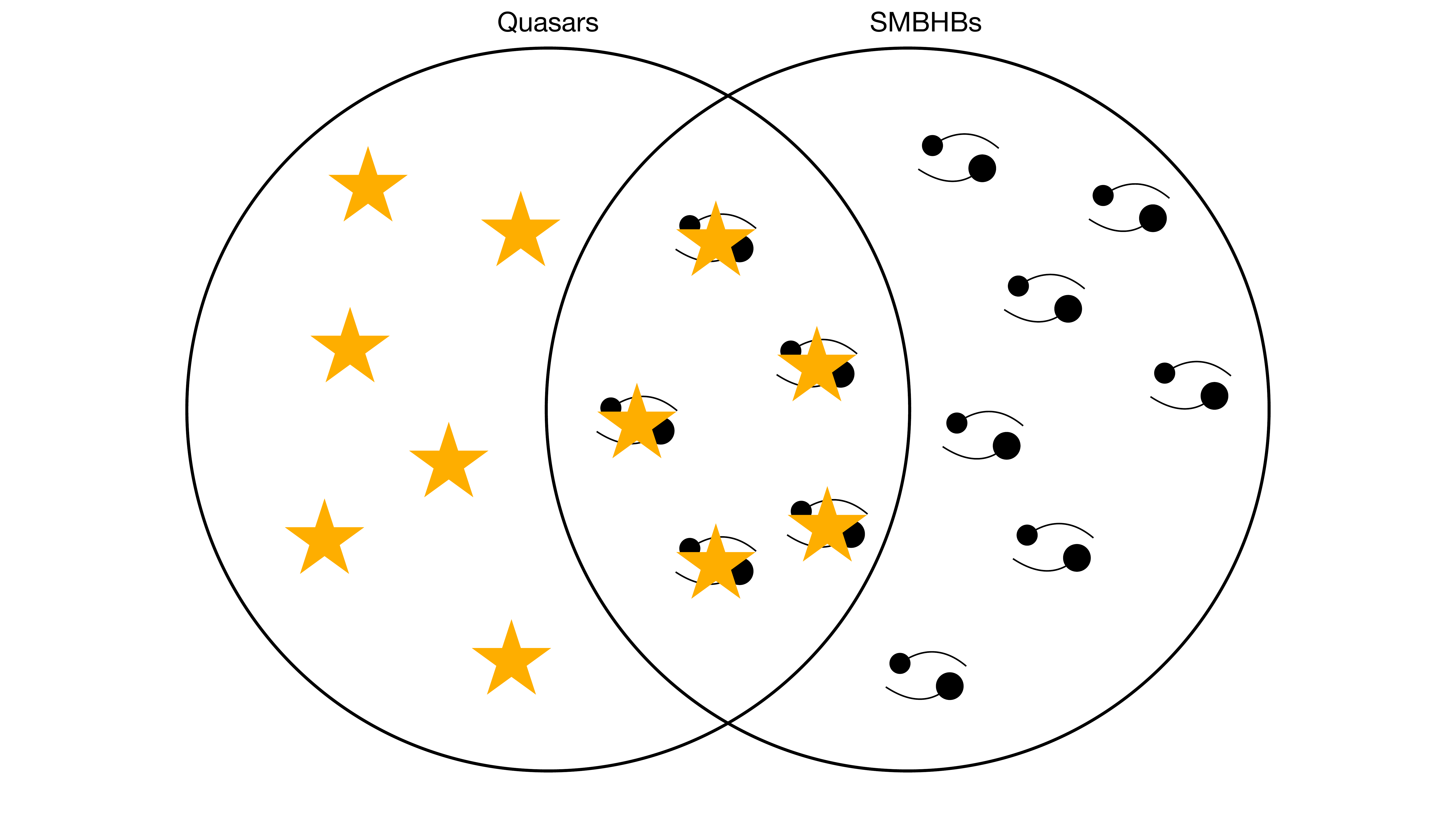}
        \caption{Visualization of the quasar (gold stars, left circle) and SMBHB populations (cartoon binaries, right circle), and the potential overlap of these populations (middle) as SMBHBs with associated quasar activity.}
        \label{fig:venn}
    \end{figure}
    
    Consider a population of quasars with number density $\Phi_{\rm{Q}, z}(z)$, shown as the left circle in \Cref{fig:venn}.
    At each $z$ a subset of these quasars may be associated with SMBHBs, shown by the overlapping region of \Cref{fig:venn}, with number density $\Phi_{\rm{BHB, Q}, z}(z)$.
    The fraction of quasars associated with merging binaries is then
    \begin{equation}
    \label{eq:frac_bhb_given_agn}
        f_{\rm{BHB}}(z|\rm{Quasar}) = \frac{\Phi_{\rm{BHB, Q}, z}(z)}{\Phi_{\rm{Q}, z}(z)}.
    \end{equation}
    Next consider the population of SMBHBs with number density $\Phi_{\rm{BHB}, z}(z)$, i.e. the right circle in \Cref{fig:venn}. Similar to \Cref{eq:frac_bhb_given_agn}, the fraction of SMBHBs associated with quasars is
    \begin{equation}
    \label{eq:frac_agn_given_bhb}
        f_{\rm{Q}}(z|\rm{BHB}) = \frac{\Phi_{\rm{BHB, Q}, z}(z)}{\Phi_{\rm{BHB}, z}(z)}.
    \end{equation}
    Taking the ratio of \cref{eq:frac_bhb_given_agn,eq:frac_agn_given_bhb}, we then have
    \begin{equation}
    \label{eq:ratio}
        \frac{f_{\rm{BHB}}(z|\rm{Quasar})}{f_{\rm{Q}}(z|\rm{BHB})} = \frac{\Phi_{\rm{BHB}, z}(z)}{\Phi_{\rm{Q}, z}(z)},
    \end{equation}
    which locally is
    \begin{equation}
    \label{eq:local_ratio}
        \frac{f_{\rm{BHB}}(z=0|\rm{Quasar})}{f_{\rm{Q}}(z=0|\rm{BHB})} = \frac{\Phi_{\rm{BHB}, 0}}{\Phi_{\rm{Q}, 0}},
    \end{equation}
    matching \Cref{eq:duty_cycle_frac}.
    Importantly, \Cref{eq:ratio} can be applied generally to any combination of quasar and SMBHB population models.
    We are limited here to an analysis of \Cref{eq:local_ratio} only by the construction of our SMBHB model.

\end{document}